\documentclass[12pt,preprint]{emulateapj}

\usepackage{graphicx}

\shorttitle{Phase Variations of Kepler-11}
\shortauthors{Dawn M. Gelino \& Stephen R. Kane}
\slugcomment{Submitted for publication in the Astrophysical Journal}

\begin{document}

\title{Phase Curves of the Kepler-11 Multi-Planet System}
\author{Dawn M. Gelino}
\affiliation{NASA Exoplanet Science Institute, Caltech, MS 100-22, 770 South Wilson Avenue, Pasadena, CA 91125, USA}
\email{dawn@ipac.caltech.edu}
\author{Stephen R. Kane}
\affiliation{Department of Physics and Astronomy, San Francisco State University, 1600 Holloway Avenue, San Francisco, CA 94132, USA}


\begin{abstract}

The Kepler mission has allowed the detection of numerous multi-planet exosystems where the planetary orbits are relatively compact. The first such system detected was Kepler-11 which has six known planets at the present time. These kinds of systems offer unique opportunities to study constraints on planetary albedos by taking advantage of both the precision timing and photometry provided by Kepler data to monitor possible phase variations. Here we present a case study of the Kepler-11 system in which we investigate the phase modulation of the system as the planets orbit the host star. We provide predictions of maximum phase modulation where the planets are simultaneously close to superior conjunction. We use corrected Kepler data for Q1-Q17 to determine the significance of these phase peaks. We find that data quarters where maximum phase peaks occur are better fit by a phase model than a ``null hypothesis'' model.

\end{abstract}



\section{Introduction}

The field of exoplanets is rapidly evolving.  We have progressed from simply finding new planets to characterizing them.  As of 2014 March 10, the NASA Exoplanet Archive\footnote{\tt http://exoplanetarchive.ipac.caltech.edu}\citep{ake13} reports 1,692 planets confirmed around 1,024 stars.  Additionally, NASA's {\it Kepler} mission has announced several thousand more likely transiting exoplanet candidates \citep{bor11a,bor11b,bat13, bur14}.  The abundance of high signal-to-noise data from {\it Kepler} is allowing us to obtain planetary radii measurements which facilitate characterization studies of planetary densities and therefore planetary interiors \citep{lop13,for13}.  The exquisite data also allows for other forms of study such as detection of planetary signatures from phase variations as they orbit their host star.  A few planets have been the subject of phase variation studies, including HAT-P-7b \citep{bor09,wel10}, Kepler-10b \citep{bat11}, Kepler-7b \citep{dem11}, TrES-2b \citep{kip11,bar12}, and Kepler-41b \citep{qui13}. The phase variations of an exoplanet are caused by the observed reflected light component of an exoplanet as it orbits the host star and changes phase.  The first observations of phase variations \citep{har06,knu07} followed closely after measurements of secondary eclipses were used to infer the temperatures of the planets \citep{cha05,dem05}. Infrared exoplanetary phase curves can help to map the energy redistribution of the planet \citep{knu09a,knu09b} while optical phase curves provide insight into the scattering properties of an exoplanet's atmosphere \citep{sud05,bur08}.

Along with the growing number of planets and planet candidates, the number of exoplanet systems with multiple planets has risen to almost 500. The advent of precise data from Kepler has brought about the opportunity to simultaneously observe the phase variations of these multi-planet systems. These systems offer unique opportunities to measure albedos thanks to the precision of not only the {\it Kepler} photometry, but also its timing measurements, which can accurately predict the times when maximum phase amplitude for each of planets in the system should occur. Detailed measurements of phase variations can significantly contribute to current theoretical models of exoplanet atmospheres. We have examined in detail the dependence of phase curves on eccentricity in \citet{kag10}, and \citet{kag11} examined the dependence on inclination. In addition, \citet{kag13} have developed techniques for decoupling the phase variations of planets in these multi-planet systems \citet{mad12} have created a technique which can be used to interpret phase curves as a function of orbital parameters and atmospheric reflective properties (Lambert versus Rayleigh, etc.). Other hypotheses have also been empirically derived. For example, based on a study of 24 planets with available secondary eclipse and phase variation constraints, \citet{cow11} suggest that very hot giant exoplanets may have low heat redistribution efficiency, while ``cooler'' hot Jupiters may have a variety of redistribution efficiencies.

\begin{deluxetable*}{cccccccc}
\tablecolumns{8}
\tablewidth{0pc}
\tablecaption{\label{planchar} Planetary Orbital Parameters and Derived Characteristics}
\tablehead{
  \colhead{Planet} &
  \colhead{$P\,^{\dagger}$} &
  \colhead{$T_0\,^{\dagger}$} &
  \colhead{$M_p\,^{\ddagger}$} &
  \colhead{$R_p\,^{\ddagger}$} &
  \colhead{$\rho_p\,^{\ddagger}$} &
  \colhead{$a\,^{\ddagger}$} &
  \colhead{$i\,^{*}$}  \\
  \colhead{} &
  \colhead{(days)} &
  \colhead{(date)} &
  \colhead{($M_\oplus$)} &
  \colhead{($R_\oplus$)} &
  \colhead{($g\,cm^{-3}$)} &
  \colhead{(AU)} &
  \colhead{(deg)} 
}

\startdata
b & $10.3039^{+0.0006}_{-0.0010}$ & $689.7378^{+0.0026}_{-0.0047}$ & $1.9^{+1.4}_{-1.0}$ & $1.80^{+0.03}_{-0.05}$  & $1.72^{+1.25}_{-0.91}$  & $0.091^{+0.001}_{-0.001}$ & $89.64^{+0.36}_{-0.18}$ \\

c & $13.0241^{+0.0013}_{-0.0008}$ & $683.3494^{+0.0014}_{-0.0019}$ & $2.9^{+2.9}_{-1.6}$ & $2.87^{+0.05}_{-0.06}$ & $0.66^{+0.66}_{-0.35}$  & $0.107^{+0.001}_{-0.001}$ & $89.59^{+0.41}_{-0.16}$ \\

d & $22.6845^{+0.0009}_{-0.0009}$ & $694.0069^{+0.0022}_{-0.0014}$ & $7.3^{+0.8}_{-1.5}$ & $3.12^{+0.06}_{-0.07}$ & $1.28^{+0.14}_{-0.27}$   & $0.155^{+0.001}_{-0.001}$ & $89.67^{+0.13}_{-0.16}$\\

e & $31.9996^{+0.0008}_{-0.0012}$ & $695.0755^{+0.0015}_{-0.0009}$ & $8.0^{+1.5}_{-2.1}$ & $4.19^{+0.07}_{-0.09}$ & $0.58^{+0.11}_{-0.16}$   & $0.195^{+0.002}_{-0.002}$ & $88.89^{+0.02}_{-0.02}$ \\

f & $46.6888^{+0.0027}_{-0.0032}$ & $718.2710^{+0.0041}_{-0.0038}$ & $2.0^{+0.8}_{-0.9}$  & $2.49^{+0.04}_{-0.07}$ & $0.69^{+0.29}_{-0.32}$   & $0.250^{+0.002}_{-0.002}$ & $89.47^{+0.04}_{-0.04}$ \\

g & $118.3807^{+0.0010}_{-0.0006}$ & $693.8021^{+0.0030}_{-0.0021}$ & $<25$              & $3.33^{+0.06}_{-0.08}$ & $<4$ & $0.466^{+0.004}_{-0.004}$ & $89.87^{+0.05}_{-0.06}$ 
\enddata

\tablenotetext{$\dagger$}{From Table 1 of \citet{lis13}.}
\tablenotetext{$\ddagger$}{From Table 4 of \citet{lis13}.}
\tablenotetext{*}{From Table 2 of \citet{lis13}.}
\end{deluxetable*}

The Kepler-11 multi-planet system is one of the earliest discovered of the {\it Kepler} systems \citep{lis11} and has been studied and characterized in sufficient detail to greatly improve the phase model.  Despite having a {\it Kepler} magnitude of 13.709 (NASA Exoplanet Archive), which places it midway between the brightest and faintest of the {\it  Kepler} systems, Kepler-11 represents an idealized case of a compact multi-planet system comprised of relatively large, in a Kepler sense, planets which should produce the maximum cumulative flux ratio when all planets are near superior conjunction. Also, the planets in this system fall into a radius regime where the geometric albedos are largely unknown.  Therefore, in this paper we investigate the phase variations of the tightly packed Kepler-11 mutli-planet system in an effort to constrain its planetary albedos.  In Section 2, we present the characteristics of the overall system which are input in to the system's flux ratio model.  We use an atmosphere model to calculate the system phase variations in Section 3 and also show the system configuration at times of peak flux amplitude. In Section 4, we describe the processing of the long cadence Kepler data.  We present our results from fitting and extracting phase signatures for the Kepler-11 system along with our subsequent constraints for the planetary albedos in Section 5.


\section{Kepler-11 System Characteristics}

The Kepler-11 system is comprised of six known transiting planets orbiting a slightly evolved $T_\mathrm{eff} \sim 5660$ K star with $M_{*} \sim 0.96 M_{\odot}$ and $R_{*} \sim 1.07 R_{\odot}$ \citep{lis13}. The orbital parameters for the six planets in the Kepler-11 system (orbital period, $P$, time of mid-transit, $T_{0}$, mass, $M_{p}$, radius, $R_{p}$, density, $\rho_{p}$, semi-major axis, $a$, and orbital inclination angle, $i$) are presented in Table \ref{planchar}.  These parameters, taken from \citet{lis13}, are used as input to the phase curve models presented in Section \ref{phasevar}.



\subsection{Planetary Properties}

The planetary orbital periods in this system are tightly packed with five that are less than 50 days, and the remaining, outermost planet, $g$, at $\sim$118 days. All of the orbits are roughly within 1\degr of being edge-on to our line of sight. Also, as with many other Kepler multi-planet systems \citep{lis12,bor13}, the eccentricities of the planets are reported to be very small and/or consistent with a circular orbit \citep{lis13}.

The planets themselves have masses and radii that range from $\sim$ 2 to 8 $M_{\oplus}$, and $\sim$ 2 to 4 $R_{\oplus}$ respectively.  This implies that these planets are very low-density and must be comprised of large amounts of very light elements.  Based on work by \citet{lop12} and \citet{lis13}, five out of the six planets in the Kepler-11 system are found to have H/He envelopes that account for about half of each of their observed radii.  The sixth planet's H/He envelope is still significant, but only accounts for roughly one third of its radius.  These large gaseous envelopes will help constrain the determination of the planetary albedos and phase curve modeling parameters.


\subsection{Geometric Albedos}

As noted in Table \ref{planchar}, the planets in this system all have radii ranging from two to four times that of the Earth, which classifies them as Neptune- and mini-Neptune-sized.  The geometric albedos of giant planets in this radius regime are largely unknown, but are thought to be dependent on the location and surface conditions of the planet. \citet{kag10} quantified the theoretical models of \citet{sud05} that showed a dependence of geometric albedo of gas giant planets on the semi-major axis of the system. The star--planet separation has an effect on the removal of reflective condensates in the upper atmospheres of the planets \citep{sud05,bur08}, and since Neptune-sized planets potentially have a more diverse atmosphere composition than their larger counterparts, there is likely a greater diversity in their albedos \citep{mos13,mig14}.

The densities of the planets in this system (probably including planet g) range from that of Neptune (at 1.76) to one third of that value. This suggests that they can be safely treated as gas giants. As noted by \citet{lis13}, approximately half of the radii of planets c, d, e, and f are due to their H/He envelopes.  Even planet b is estimated to have a rocky core that comprises only 66\% of its radius.  We assume a proxy albedo in this largely unexplored region of albedo space, and we will use the data fits to our models to constrain this parameter.


\section{Phase Variations}

\label{phasevar}
In this section we model the phase variations of the six-planet Kepler-11 system.  We follow the model detailed in \citet{kag10} for the phase variations and geometric albedo of a planet.  For each planet, the phase angle $\alpha$ is defined to be zero at superior conjunction and is described by

\begin{equation}
  \cos \alpha = \sin (\omega + f) \sin i
  \label{phaseangle}
\end{equation}

where $f$ is the true anomaly and $i$ is the inclination of the orbit. The phase function of a Lambert sphere assumes isotropic scattering of incident flux over $2 \pi$ steradians and is determined by

\begin{equation}
  g(\alpha,\lambda) = \frac{\sin \alpha + (\pi - \alpha) \cos \alpha}{\pi}
\end{equation}

This formalism is used for rocky and molten surface models \citep{kan11}.  For the atmosphere model used below, we adopt the phase function of \citet{hil92} which is empirically derived based on observations of Jupiter and Venus. This model contains a correction to the planetary visual magnitude

\begin{equation}
  \Delta m (\alpha) = 0.09 (\alpha/100\degr) + 2.39 (\alpha/100\degr)^2 -0.65 (\alpha/100\degr)^3
\end{equation}

which leads to a phase function given by

\begin{equation}
  g(\alpha) = 10^{-0.4 \Delta m (\alpha)}.
\end{equation}

This form allows for non-isotropic (cloud) scattering. 

At phase angle $\alpha = 0\degr$, the geometric albedo of a planet is defined as:

\begin{equation}
  A_g(\lambda) = \frac{F_r(0,\lambda)}{F_i(\lambda)}
\end{equation}

where $F_r (\lambda)$ is the reflected light from the planet at wavelength $\lambda$, and similarly, $F_i (\lambda)$ is the incident flux on the planet, which is defined as:

\begin{equation}
  F_i(\lambda) = \frac{L_\star(\lambda)}{4 \pi r^2}
\end{equation}

where $L_\star$ is the luminosity of the star and $r$ is the star--planet separation.  \citet{sud05} studied flux ratio dependencies on wavelength.  Our study here is limited to optical wavelengths centered on 550~nm, near the peak response of the {\it Kepler} detectors. Table \ref{albedoflux} shows our planetary albedos for each of the Kepler-11 planets determined in this way.

Finally, the flux ratio of the planet to the host star is defined as

\begin{equation}
  \epsilon(\alpha,\lambda) \equiv \frac{f_p(\alpha,\lambda)}{f_\star(\lambda)} = A_g(\lambda) g(\alpha,\lambda) \frac{R_p^2}{r^2}
  \label{fluxratio}
\end{equation}

where $R_p$ is the planetary radius and $r$ is the star--planet separation. For a circular orbit, only the phase function is time dependent.


\subsection{Kepler-11 Phase Modulation}

\label{phasemodulation}

Using our formulation for planetary phase variations, we construct a complete model for the expected phase modulations in the Kepler-11 system during the times of observations. We utilize Kepler data spanning Q1 through Q17 covering a total time span of 1460 days (see Section \ref{dataprep}). Our calculations for the geometric albedo (dependent on star--planet separation) for each planet are shown in Table \ref{albedoflux} along with their peak phase amplitudes.


\begin{deluxetable}{ccc}
\tablecolumns{3}
\tablewidth{0pc}
\tablecaption{\label{albedoflux} Derived Planetary Albedos and Flux Ratios}
\tablehead{
  \colhead{Planet} &
  \colhead{$A_g$} &
  \colhead{Flux Ratio (10$^{-6}$)}
}
\startdata
b & 0.156 & 0.11 \\
c & 0.157 & 0.21 \\
d & 0.162 & 0.12 \\
e & 0.167 & 0.14 \\
f & 0.173 & 0.03 \\
g & 0.202 & 0.02
\enddata
\end{deluxetable}

\begin{figure*}
  \begin{center}
    \includegraphics[angle=270,width=15.0cm]{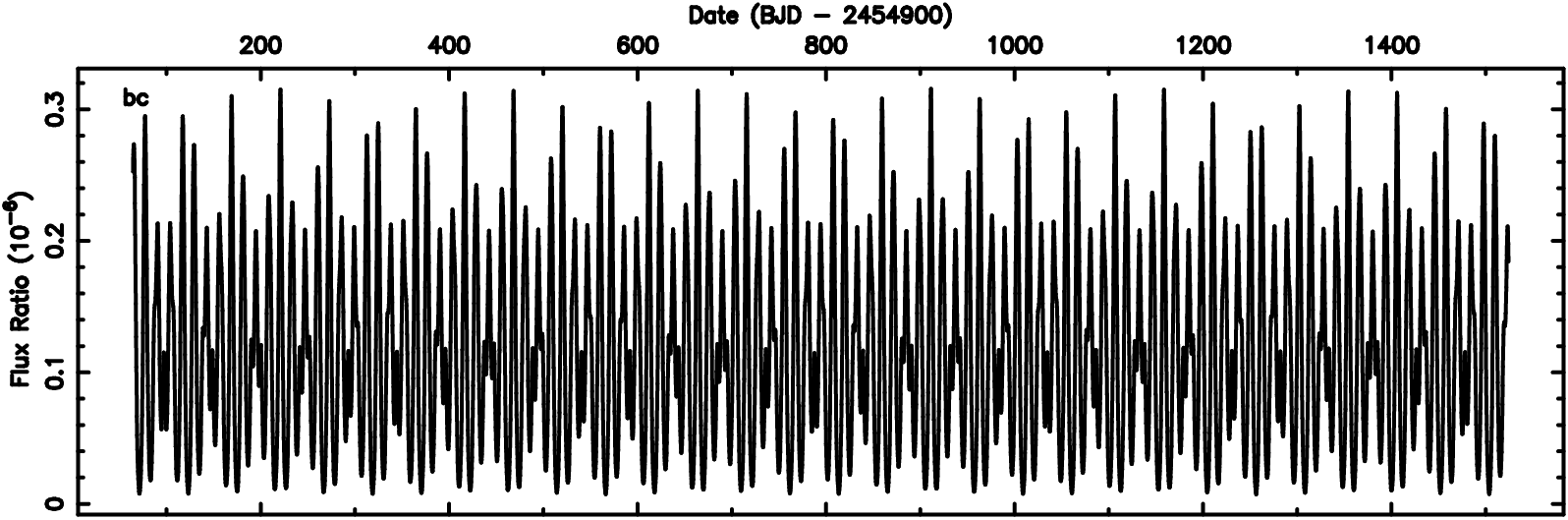} \\
    \includegraphics[angle=270,width=15.0cm]{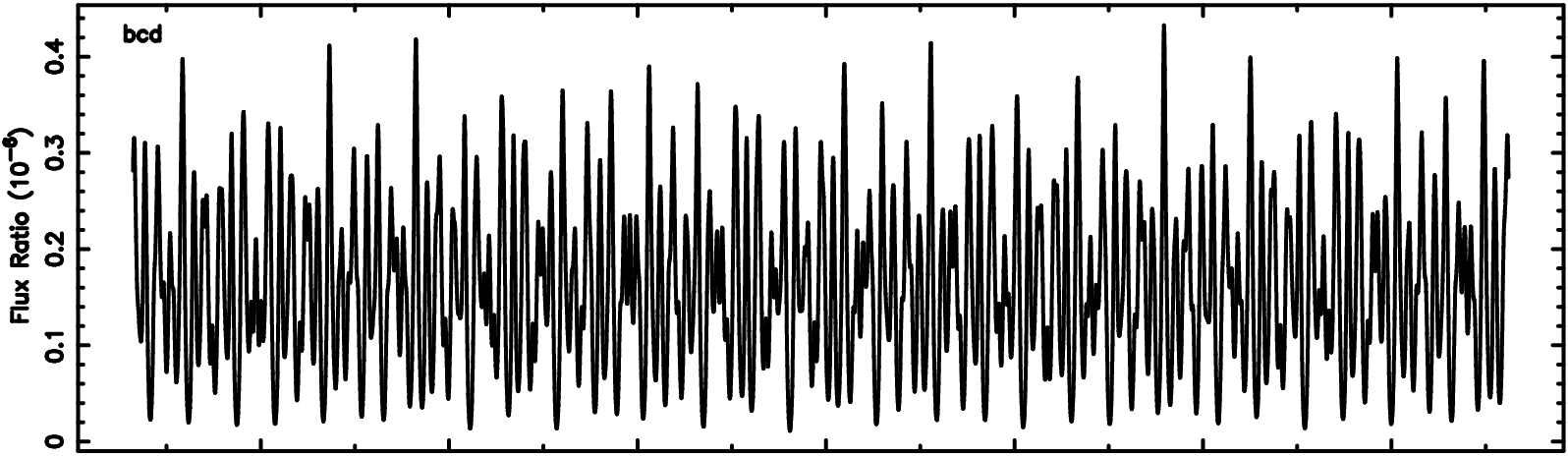} \\
    \includegraphics[angle=270,width=15.0cm]{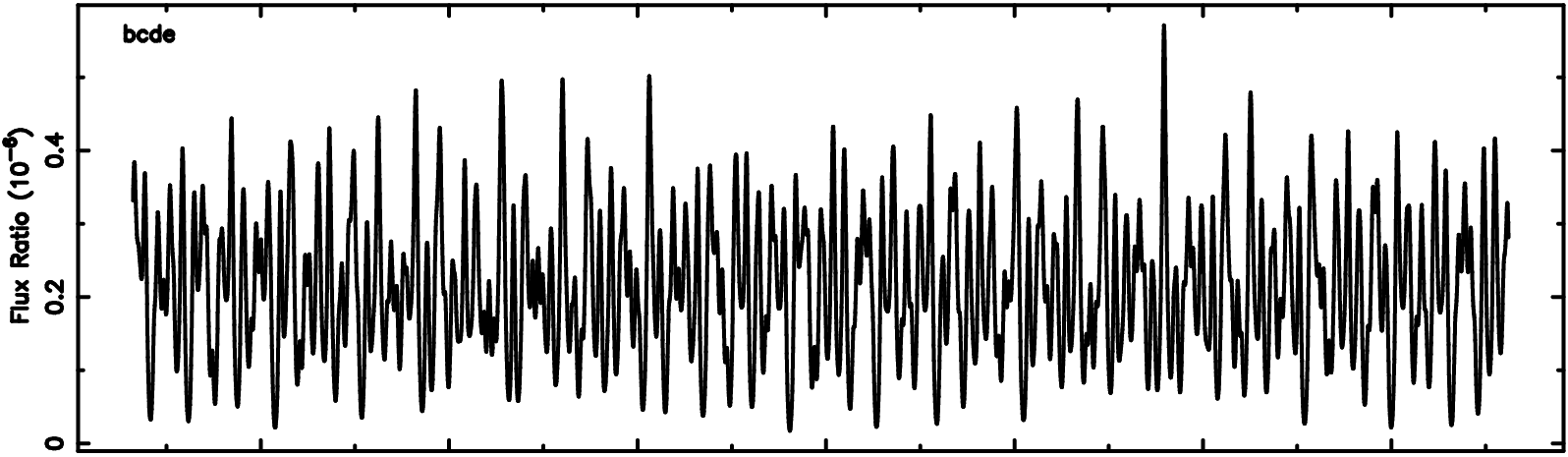} \\
    \includegraphics[angle=270,width=15.0cm]{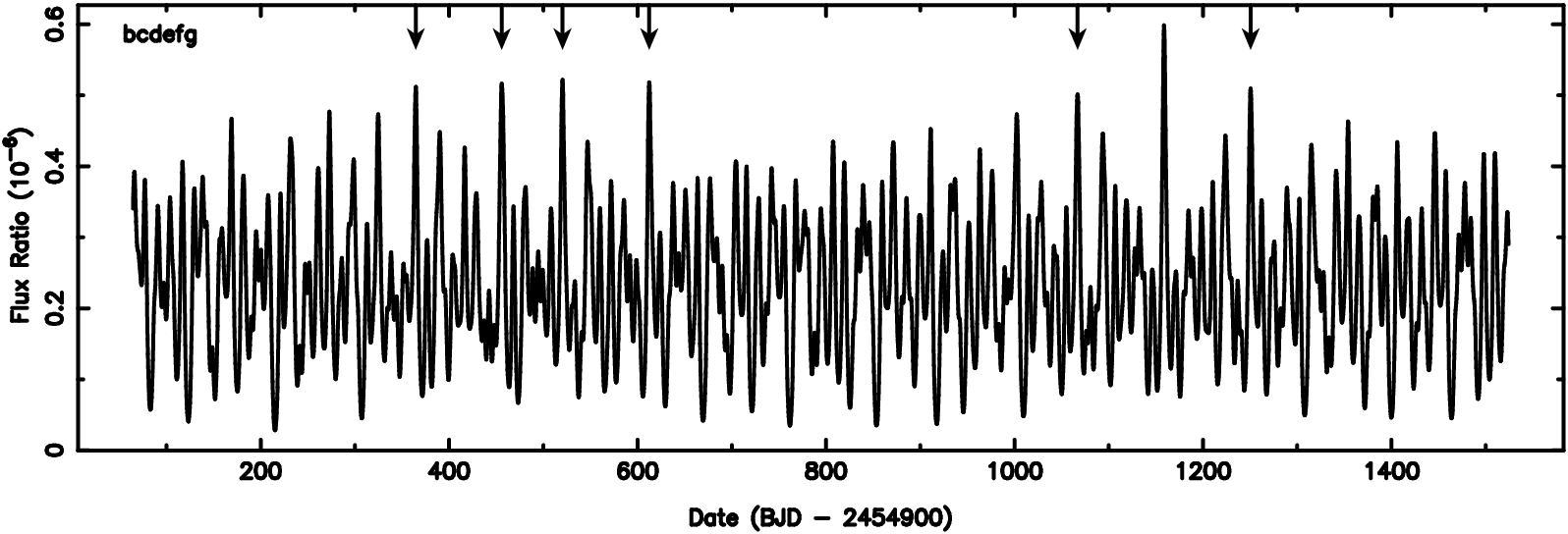}
  \end{center}

  \caption{Predicted flux variations of the Kepler-11 system computed over the timespan of Q1--Q17 Kepler observations. The calculations assume an atmosphere model for the planets. Each of the panels includes the flux contributions of a different set of planets, indicated by the planet letter designations shown in the top-left of the panel. The lower panel's similarity to the panel above it shows that the flux contribution of the two outer (fg) planets is negligible when combined with that of the four inner (bcde) planets. The arrows in the lower panel indicate flux peaks which exceed $0.5 \times 10^{-6}$ and will be the subject of further analysis in Section \ref{results}.}
  \label{modelphase}
\end{figure*}

Each of the planets contribute to the total phase amplitude in different ways at different times. To demonstrate their relative contributions to the phase modulation, we show in Figure \ref{modelphase} the predicted modulation when including different combinations of the planets. The planets included for the models are shown in the top-left of each panel. The inner four (bcde) planets are the major contributors to the modulation and the shape of the modulation changes significantly between each of these combinations. The outer two (f and g) planets are minor contributors and have a minor effect on the modulations. The figure also shows the times of peak amplitude when most of the planets are close to superior conjunction. There are seven primary peaks with amplitudes greater than $0.5 \times 10^{-6}$ which occur at times of 364.59 (Q4), 455.82 (Q5), 520.28 (Q6), 612.38 (Q7), 1066.92 (Q12), 1158.72 (Q13), and 1250.59 (Q14) where times are expressed as BJD $- 2454900$. These peaks are indicated by arrows in the lower panel of Figure \ref{modelphase}. Unfortunately the predicted peak during Q13 occurred at a time of significant stellar activity so we do not consider this peak in the subsequent analysis. Section \ref{results} presents the results of fitting the Kepler data to the remaining six times of maximum phase modulation.


\subsection{Orbital Configurations}
\label{orbconfig}

The precise timing of the Kepler-11 planets (see Table \ref{planchar}) allows us to construct a detailed orrery of the system. As described earlier, we are primarily investigating times of maximum observed planetary fluxes when most of the planets are near superior conjunction. We can thus test the location of the planets at the predicted times of maximum flux (see Section \ref{phasemodulation}) to visualize how each are contributing to the phase variations at those times.

\begin{figure*}
  \begin{center}
    \begin{tabular}{ccc}
      \includegraphics[angle=270,width=5.5cm]{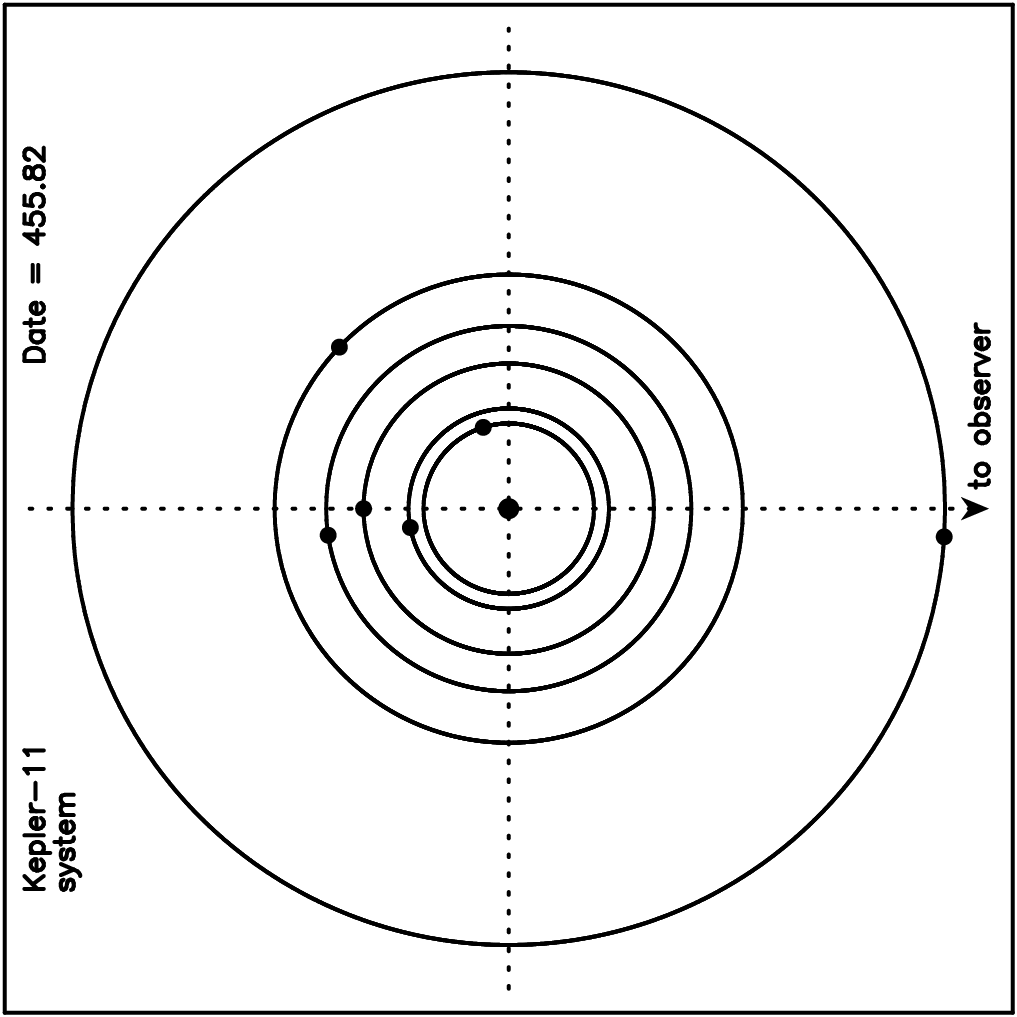} &
      \includegraphics[angle=270,width=5.5cm]{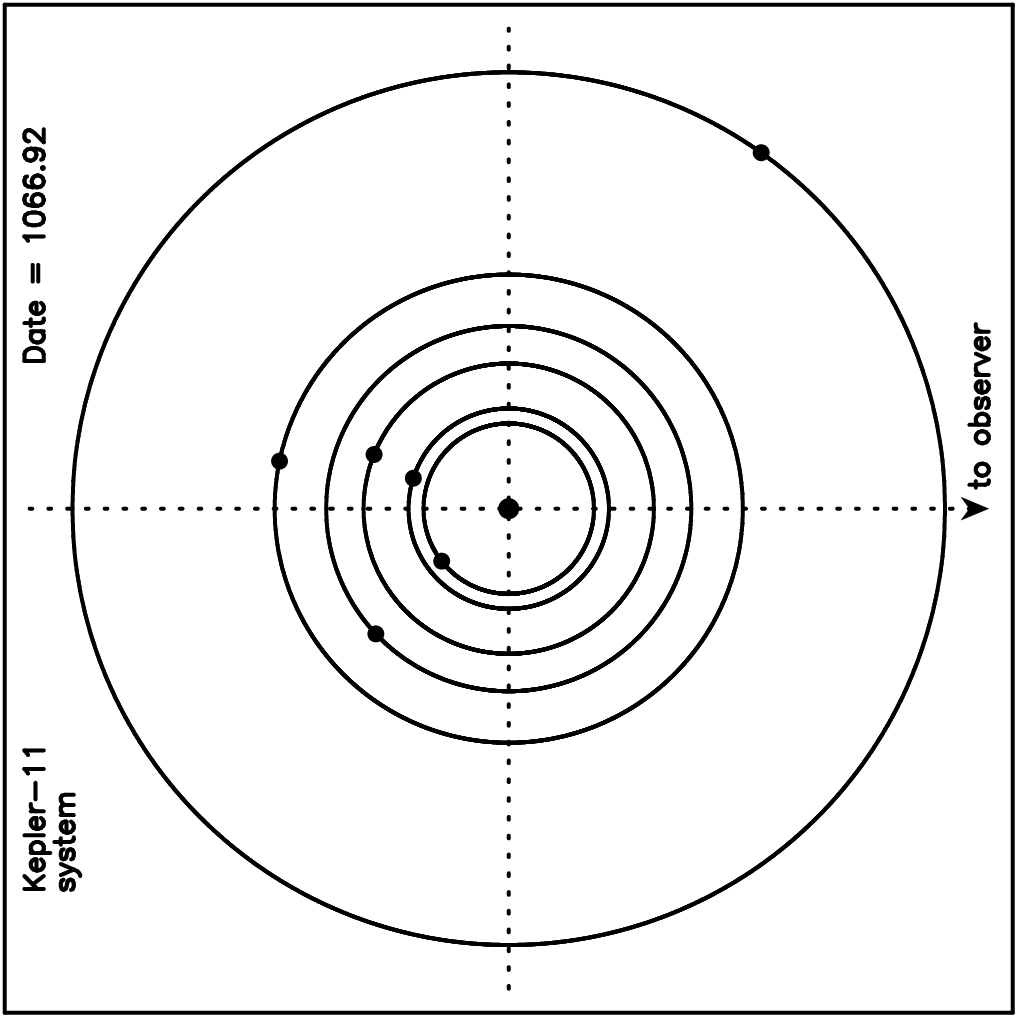} &
      \includegraphics[angle=270,width=5.5cm]{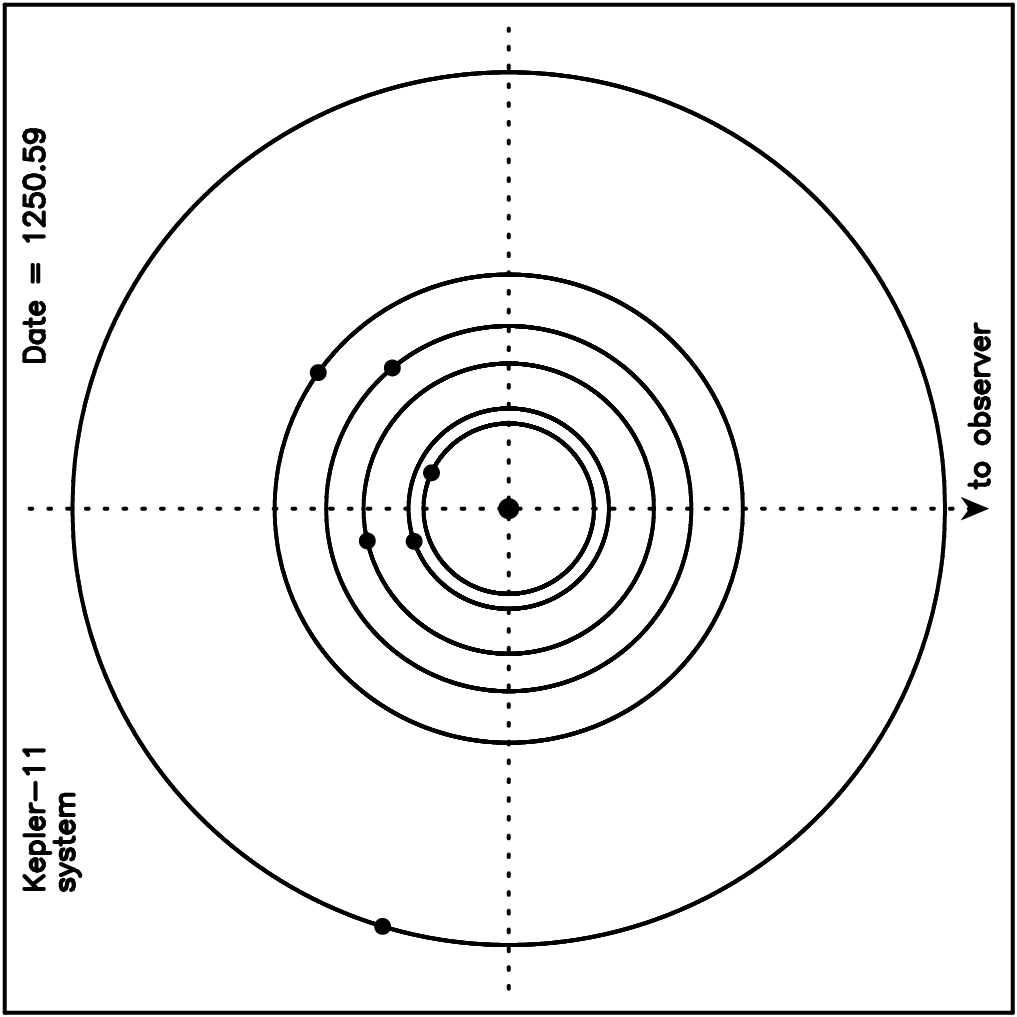}
    \end{tabular}
  \end{center}
  \caption{Top-down views of the Kepler-11 system which depict the location of the planets with respect to the observer line of sight at the times of peak phase amplitude. The three orbital configurations depicted here correspond to the phase peaks at times 455.82, 1066.92, and 1250.59 (see Section \ref{phasemodulation}). These plots confirm that the peak phase amplitude for the system occur when most of the planets are near superior conjunction.}
  \label{orbconfigplot}
\end{figure*}

Shown in Figure \ref{orbconfigplot} are top-down views of the Kepler-11 system for three of the main peaks described in Section \ref{phasemodulation} and Figure \ref{modelphase}. In each case, we show the orientation of the system with respect to the observer and the location of the six planets at the time of observation. These plots further demonstrate the lack of flux contribution from the outer two planets since their orbital locations can be far from superior conjunction at the time of maximum phase amplitude.


\section{Data Preparation}
\label{dataprep}

\begin{figure*}
  \begin{center}
    \includegraphics[angle=270,width=15.0cm]{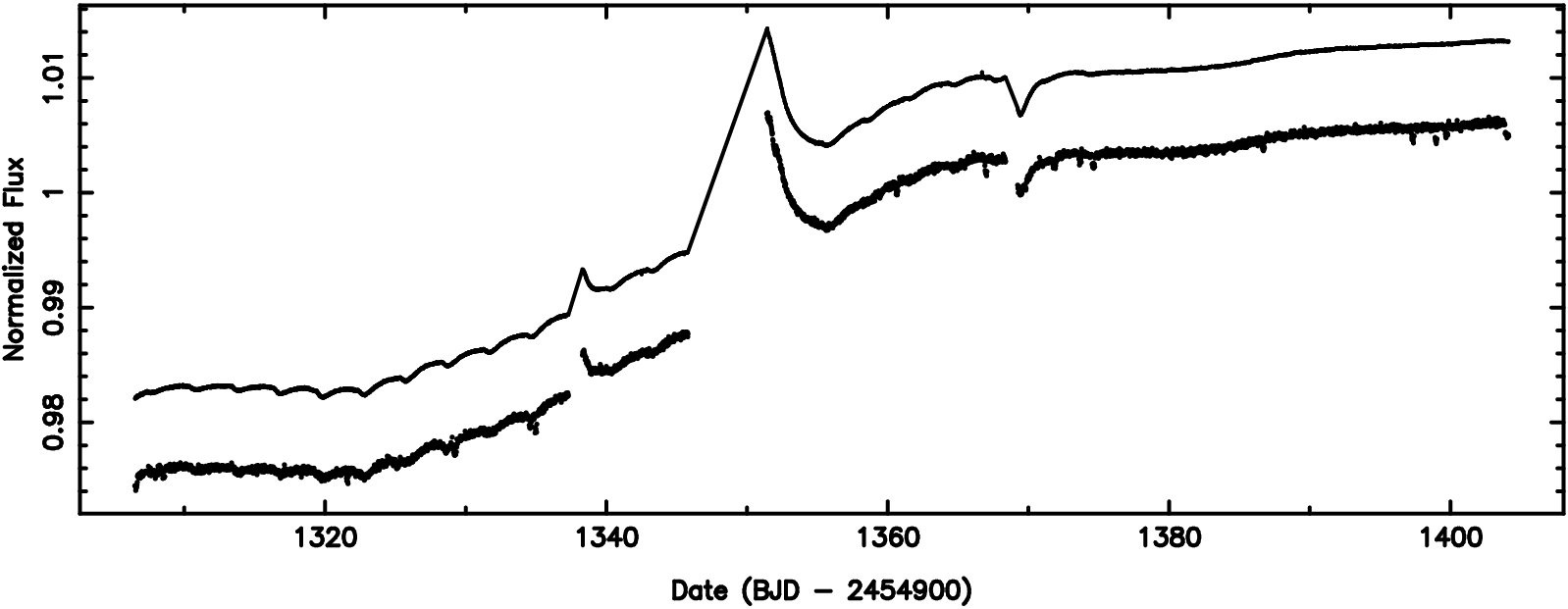} \\
    \includegraphics[angle=270,width=15.0cm]{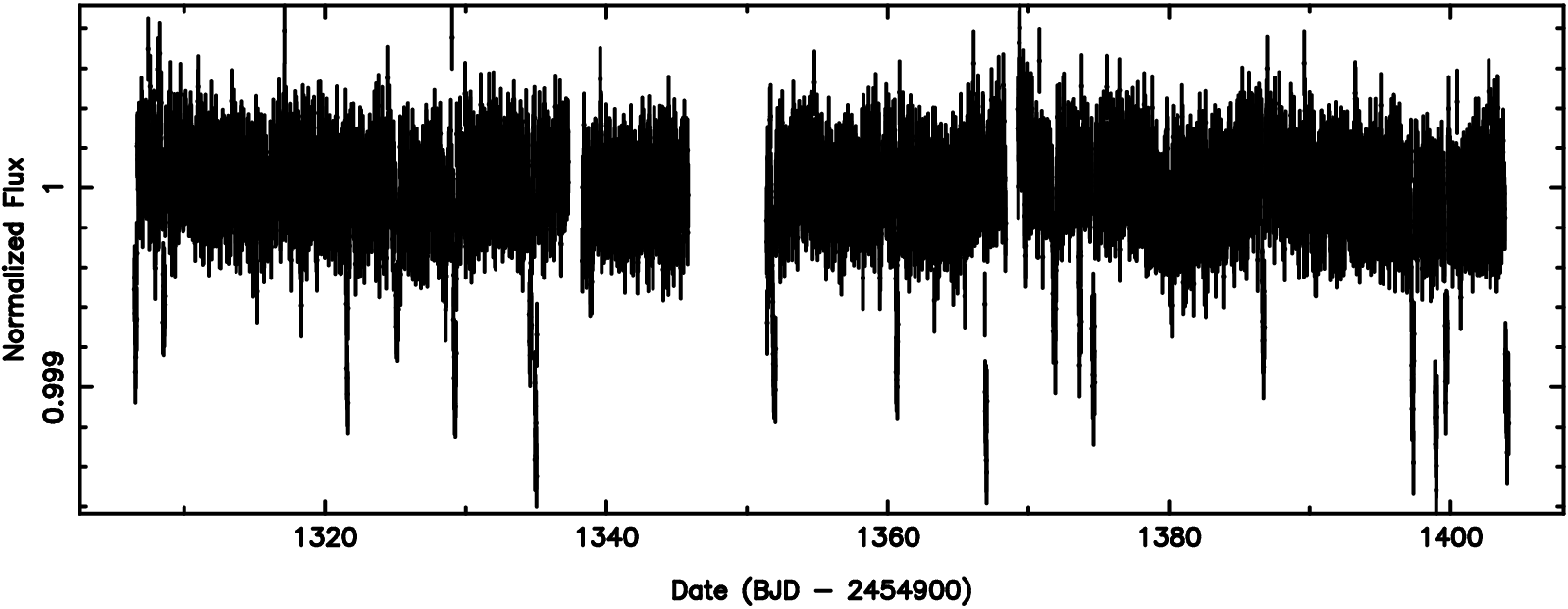}
  \end{center}

  \caption{Kepler-11 photometry for Q15. Top panel: Raw (SAP) photometry along with the best-fit model produced using the Q15 CBVs. Bottom panel: Corrected normalized flux for Q15 where the CBV model has been applied to the SAP flux.}
  \label{dataprepplot}
\end{figure*}

The Q1--Q17 public data for Kepler-11 were extracted from the Mikulski Archive for Space Telescopes (MAST)\footnote{\tt http://archive.stsci.edu/kepler/}. These data are current as of 2013 December 9. We use only the long cadence data for each quarter since this data provide significantly better photometric precision than the high cadence data, and high cadence data is not required to adequately sample the orbital phase of the planet. Given this, the higher precision and lower cadence of the data is better suited to the phase analysis described here. We used the cotrending basis vectors (CBVs) that were part of the same data release. These data provide simple aperture photometry (SAP) fluxes which may be manipulated using a combination of the CBVs to account for instrumental noise sources without compromising variations that are astrophysical in origin.

Shown in Figure \ref{dataprepplot} are the Q15 data as an example of our data processing. The top panel shows the SAP photometry which have been normalized by the data median. The solid line is the best-fit model to the instrumental noise produced using the first two CBVs for Q15. Applying this model results in the normalized data shown in the bottom panel. In this plot we have included the data uncertainties which shows that they are consistent with the RMS scatter in the data. For this particular quarter, there are minor variations which remain in the time series. These are inconsequential to our phase analysis since we examine epochs of maximum phase amplitude, as described in Section \ref{phasevar}.

Two further steps were taken to prepare the data for analysis. The transit signatures present in the data were removed by applying a 2$\sigma$ clip. In general, this removed only $\sim 3$\% of the total number of measurements in each quarter. The data were also binned in intervals of 6 hours. This reduced the number of measurements by factor of 12 and the 1$\sigma$ scatter by a factor of 2 for each quarter. The result is a 1$\sigma$ scatter in the relative flux of $\sim 8.0 \times 10^{-5}$ for a typical quarter.


\section{Extracting Phase Signatures}

\label{results}

Here we describe the phase analysis of the Kepler data. We examined each quarter individually for predicted phase signatures shown in Figure \ref{modelphase} with the exception of Q13 (see Section \ref{phasemodulation}). With the data preparation described in Section \ref{dataprep}, the typical phase variations are a factor of $\sim 100$ smaller than the typical 1$\sigma$ scatter in the relative flux. This is not surprising considering that the magnitude of Kepler-11 is towards the fainter end of the confirmed Kepler multi-planet system host stars.

In order to perform a meaningful test of how these data can constrain possible phase variations, we first examined the data near the times of peak phase amplitude identified in Section \ref{phasemodulation}. In each of the quarters with an identified phase peak $> 0.5 \times 10^{-6}$ we calculated the $\chi^2$ for a period of 10 days surrounding the central peak. We then compared this with a null (constant) model and calculated the $\Delta \chi^2$. For each of these quarters we obtained a positive $\Delta \chi^2$, indicating that the phase model is a slight improvement over the constant model. We further performed a similar $\chi^2$ calculation for each individual quarter and found that the $\Delta \chi^2$ values for the quarters with the large phase amplitudes were higher than those for the quarters where such planetary alignments did not produce such phase peaks. Although this is insufficient to claim a detection of the system phase variations, it demonstrates the process of such detections in the presence of very low signal-to-noise.

\begin{figure*}
  \begin{center}
    \includegraphics[angle=270,width=15.0cm]{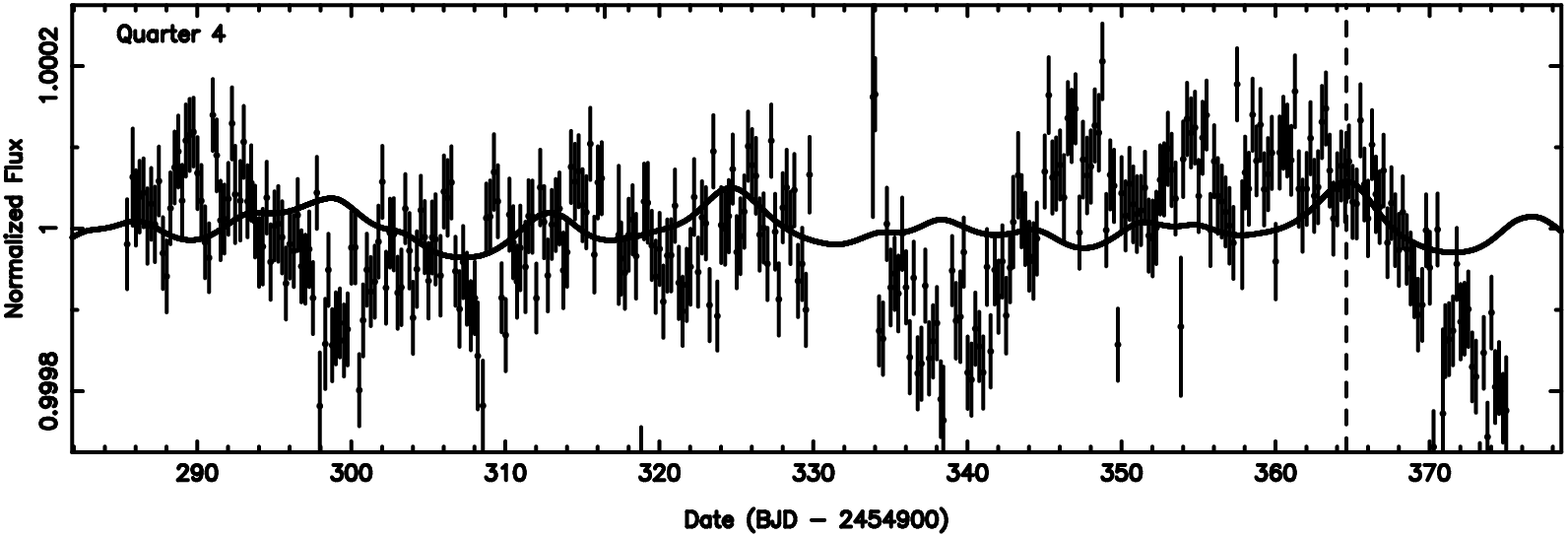} \\
    \includegraphics[angle=270,width=15.0cm]{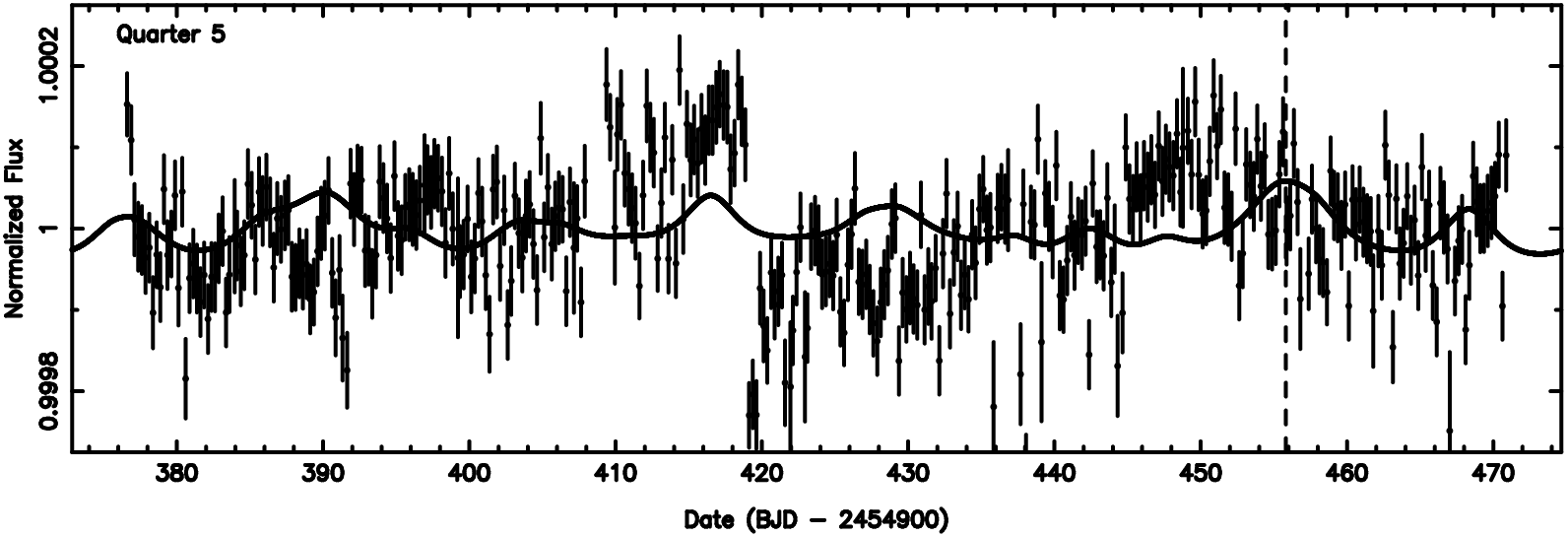} \\
    \includegraphics[angle=270,width=15.0cm]{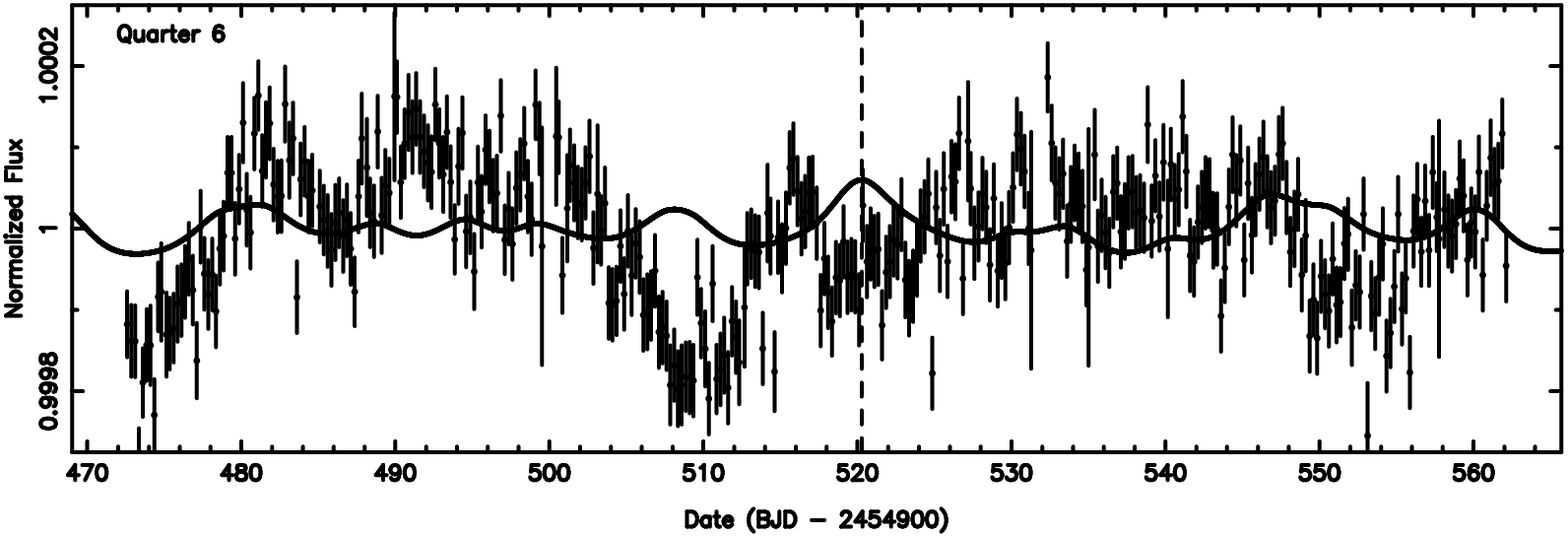} \\
    \includegraphics[angle=270,width=15.0cm]{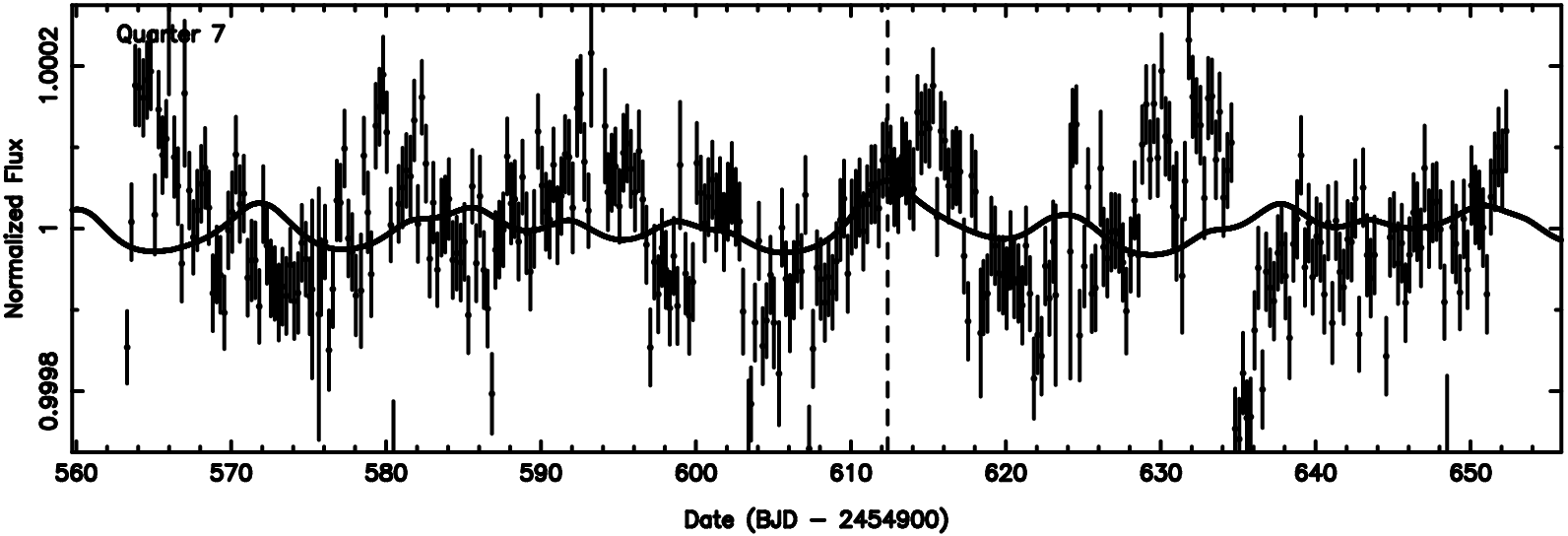}
  \end{center}

  \caption{Predicted flux variations of a modified Kepler-11 system where the radii are five times those shown in Table \ref{planchar} and the albedos are double those shown in Table \ref{albedoflux}. These are shown for Q4--Q7 along with the corresponding corrected Kepler data. The vertical dashed line in each panel shows the times of predicted maximum phase amplitude, as described in Section \ref{phasemodulation}.  Note that despite the well measured system parameters, the faintness of the Kepler-11 host star is the major contributor to the relatively large scatter in the flux data.}
  \label{phasedata}
\end{figure*}

As shown in Equation \ref{fluxratio}, the amplitude of the flux ratio depends linearly on the albedo, and on the square of the planetary radius. Thus the model is very sensitive to these two quantities and relies upon their robust determination. We tested a variety of realistic models which produce phase signatures more consistent with the noise properties of the Kepler data. One of these models is shown in Figure \ref{phasedata} in which we have increased the radii of the planets (see Table \ref{planchar}) by a factor of five such that they are now giant planets. We have also increased the geometric albedos by a factor of two compared with those values in Table \ref{albedoflux} which is more consistent with reflective condensates present in the upper atmosphere, a feature which has been detected for short-period planets \citep{dem11}. The times of peak phase amplitude (indicated by the vertical dashed lines in Figure \ref{phasedata}) are now $\sim 1.0 \times 10^{-4}$, more than enough to be detected by the Kepler data for this star. In this case, the revised system model produces an improved fit to the data for Q4 and Q7, but a worse fit for Q5 and Q6. It is worth noting that a recent study by \citet{kis13} using hydrodynamic wind models showed that the Kepler-11 planets could be experiencing significant atmospheric mass loss due to the stellar winds present within the regime of the compact system. Such mass loss effects may cause the radii of the planets to be over-estimated and the geometric albedos to be under-estimated.


\section{Discussion}

The methodology described here is a powerful means to investigate the phase variations of multi-planet systems. This is contingent upon the system parameters producing predicted phase variations of amplitude consistent with the photometric precision. In such cases, the only free parameters are the albedos of the planets which may be determined through a detection or constrained through a null detection. Section \ref{results} describes a variation of the Kepler-11 system model which increased the Table \ref{albedoflux} albedos by a factor of two.

There are a variety of models to predict the geometric albedo of a giant planet \citep{kag10,mad12,cow13} which depend greatly upon the properties of the atmosphere. Recent results from Kepler have revealed a large diversity of planetary densities in the super-Earth to Neptune size regime \citep{mar14}. This undoubtedly is accompanied by a similar diversity in planetary atmospheres and albedos. Thus it is reasonable to test models which have substantially higher albedos than those calculated from a pure gas giant perspective.

There are several factors which can influence the results of this analysis. Several studies have shown the relative prevalence of stellar activity at the precision of Kepler photometry \citep{cia11,wal13}. Thus stellar variability will often be present in the Kepler photometry at a similar amplitude to the phase signatures we are trying to extract. However, the stellar variability is not expected to be correlated with the orbital phase of the planets and will thus have a net zero effect when averaged over many quarters of data.

A source of stellar variability which is indeed induced by the presence of planets is the induction of ellipsoidal variations in the host star. An approximate relation for the amplitude of this effect by \citet{loe03} is as follows:

\begin{equation}
  \frac{\Delta F}{F_0} \sim \beta \frac{M_p}{M_\star} \left( \frac{R_\star}{a} \right)^3
  \label{ellipsoidal}
\end{equation}

where $\beta$ is the gravity darkening exponent. There are several reasons why we neglect this effect for the kinds of multi-planet systems described in this work. Firstly, the ellipsoidal variations have a $a^{-3}$ dependence compared with the $a^{-2}$ dependence of phase variations. Secondly, the amplitude depends upon planetary mass which is far more uncertain than planetary radii for these systems. Thirdly, the ellipsoidal variations achieve highest amplitude at phase angles of $90\degr$ and $270\degr$ and are thus anti-correlated with phase variations.

Similarly, Doppler boosting (beaming) can produce a similar amplitude of stellar variability as the star moves toward and away from the observer due to the presence of the planets \citep{loe03,fai11,shp11}. The fractional amplitude of the effect is as follows:

\begin{equation}
  \frac{\Delta F}{F_0} = \frac{(3-\alpha)K}{c}
  \label{boost}
\end{equation}

where $K$ is the radial velocity semi-amplitude and $\alpha$ is the derivative of the bolometric flux with respect to the frequency in the stationary frame of reference. Note that the uncertainty associated with the planetary masses poses a similar problem to Doppler boosting as it does to ellipsoidal variations. Nonetheless, the dependence on $K$ ensures that the amplitude of Doppler boosting will have a relatively small effect for low-mass planets with large orbital periods, such as the planets analyzed here.

An additional consideration is that of secondary eclipses. The Kepler multi-planet systems tend to consist of relatively well-aligned circular orbits such that one can expect secondary eclipses to regularly occur. The implication of this is that the reflected light from the planets will not be visible at a phase angle of $0\degr$ where the phase amplitude is maximum. However, here we are concerned with the times of maximum phase amplitude which occur due to the combined effect of all planets within the system. As shown in Figure \ref{orbconfigplot}, this rarely occurs when all planets are precisely at superior conjunction and so in most cases will not affect the expected outcome of our phase predictions for multi-planet systems.


\section{Conclusions}

The Kepler multi-planet systems allow accurate orreries to be constructed. One advantage of this is the prediction of observable features which coincide with specific orbital configurations. One such time-variable observable is that of phase variations. Here we have used available system properties of the compact Kepler-11 system to predict the phase modulation due to the orbital motion of the planets. By connecting these predictions to the Kepler data from Q1 to Q17, we have investigated the possibility of that signatures of peak phase amplitude may be present in the data.

Our results show that quarters with predicted maximum phase peaks, when there are a sufficient number of planets close to superior conjunction, are best fit by a phase model rather than a constant model. Although the signal-to-noise of the Kepler-11 data compared with the model does not allow this to be conclusively shown to be the cause of the correlation, it does demonstrate how this technique may be used to further investigate similar systems. We have additionally shown how the sensitivity of phase variations to planetary radius and albedo allows for a wider range of planetary systems to be explored in this way. The full list of Kepler candidates now contains many multi-planet systems for which precise timing information is available. A more thorough investigation of the phase properties of these systems will yield an additional step toward characterizing the planets contained therein.


\section*{Acknowledgements}

This research has made use of the NASA Exoplanet Archive, which is operated by the California Institute of Technology, under contract with the National Aeronautics and Space Administration under the Exoplanet Exploration Program. Some/all of the data presented in this paper were obtained from the Mikulski Archive for Space Telescopes (MAST). STScI is operated by the Association of Universities for Research in Astronomy, Inc., under NASA contract NAS5-26555. Support for MAST for non-Hubble Space Telescope data is provided by the NASA Office of Space Science via grant NNX13AC07G and by other grants and contracts. This paper includes data collected by the Kepler mission. Funding for the Kepler mission is provided by the NASA Science Mission directorate.


\end{document}